\ifx\onecol\undefined
    \documentclass[10pt,journal,twocolumn]{IEEEtran}
    \newcommand{\figsize}{0.99}
    
\else 
    \documentclass[12pt,draftclsnofoot,journal,onecolumn]{IEEEtran}
    \newcommand{\figsize}{0.7}
\fi

\IEEEoverridecommandlockouts
\usepackage{amsmath}
\usepackage{amsthm,amssymb}
\usepackage{algorithm}
\usepackage{algorithmic}
\usepackage{bm,bbm}
\usepackage{balance}
\usepackage{braket}
\usepackage{booktabs}
\usepackage{color}
\usepackage{cases}
\usepackage{eqparbox}
\usepackage{enumitem}
\usepackage{graphicx}
\usepackage{mathrsfs}
\usepackage{multirow}
\usepackage{makecell}
\usepackage{subfigure}
\usepackage{setspace}
\usepackage{threeparttable}
\usepackage{verbatim}

\usepackage[numbers,sort&compress]{natbib}

\usepackage{hyperref}
\hypersetup{hidelinks, 
colorlinks=true,
allcolors=black,
pdfstartview=Fit,
breaklinks=true}

\newcommand{\ri}{{\mathrm{i}}}

\newcommand{\diff}{{\mathrm{d}}}
\newcommand{\tr}{{\rm{tr}}}

\renewcommand{\Im}[1]{{\rm Im}{\left\{{#1}\right\}}}

\def \T {^{\mathsf{T}}}
\def \H {^{\mathsf{H}}}

\begin{document}
\title{RAQ-MIMO: MIMO for Multi-Band Rydberg Atomic Quantum Receiver } 
\author{{Jieao~Zhu,~{\textit{Student Member, IEEE}}~and~Linglong~Dai,~{\textit{Fellow,~IEEE}} }
\thanks{This work was supported in part by the National Natural Science Foundation of China (Grant No. 62325106), in part by the National Natural Science Foundation of China (Grant No. 62031019), and in part by the National Key Research and Development Program of China (Grant No. 2023YFB3811503). } 
\thanks{All authors are with the Department of Electronic Engineering, Tsinghua University, and the State Key Laboratory of Space Network and Communications, Tsinghua University, Beijing 100084, China (e-mails: zja21@mails.tsinghua.edu.cn, daill@tsinghua.edu.cn). }
\thanks{L. Dai is also with the EECS Department, Massachusetts Institute of Technology, Cambridge, MA 02139 USA. }
}

\maketitle

\begin{abstract}
  Rydberg atomic quantum receivers (RAQRs) are capable of receiving multi-band radio-frequency (RF) signals simultaneously, which are expected to break Chu's limit for classical electronic antennas. 
  However, signals from different users will interfere with each other in the optical intermediate frequency (IF) domain of the multi-band quantum receiver, which is termed the IF interference (IFI) problem.  
  To address this problem, in this paper, we propose a multi-input multi-output (MIMO) architecture for Rydberg atomic quantum receiver (RAQ-MIMO) by exploiting the additional spatial diversity of MIMO receivers.   
  Specifically, by applying the dynamic signal model of RAQRs, we clarify the physical relationship between the quantum local oscillator (LO) configurations and the multi-band gains with the concept of quantum transconductance. 
  Then, with the quantum transconductance-based signal model, we formulate the spectral efficiency (SE) maximization problem and further propose the quantum weighted minimum mean square error (qWMMSE) algorithm, which jointly optimizes the quantum LO configurations and the classical precoder/combiner matrices. 
  Furthermore, we test the qWMMSE algorithm within the standard space division multiple access (SDMA) scheme and the frequency division multiple access (FDMA) scheme. 
  Simulation results demonstrate that the qWMMSE optimization framework can significantly improve the SE of RAQ-MIMO systems for both multiple access schemes, and that RAQ-MIMO systems can outperform classical electronic receiver-based multi-user MIMO systems by eliminating the mutual coupling effect between classical antennas. 
\end{abstract}


\section{Introduction}
Quantum technologies have been recognized as a transformative paradigm that exploits quantum mechanical properties to achieve performance previously unattainable by classical technology. 
Among them, the most representative technologies are quantum computing, quantum secure communications, and quantum sensing. 
Utilizing quantum superposition, quantum computing theoretically achieves exponential speedup on some difficult computational tasks, such as breaking the Rivest-Shamir-Adleman (RSA) encryption and simulating quantum-chemical reactions~\cite{nielsen2010quantum}. Guaranteed by the quantum no-cloning theorem, the security of the key distribution in quantum communications can be ensured unconditionally~\cite{gisin2007quantum}. Exploiting the inherent sensitivity of quantum-level particles to external driving forces, quantum sensing can achieve highly precise measurements of various physical quantities, from electromagnetic fields~\cite{gordon2010quantum}, mechanical rotation/acceleration~\cite{cheiney2018navigation}, to gravity fields~\cite{janvier2022compact}. 

On the road to quantum communications and sensing, Rydberg atomic quantum receivers (RAQRs) stand out as a promising approach due to the following four advantages. 
Firstly, Rydberg atoms are alkali atoms in highly excited quantum states, where electrons in these states are extremely sensitive to external electromagnetic fields because of their large transition dipole moments. The E-field sensitivity of Rydberg atomic receivers experimentally reaches ${\rm 10\,nV/cm/\sqrt{Hz}}$~\cite{tu2024approaching}, which is comparable to the sensitivity of several ${\rm nV/cm/\sqrt{Hz}}$ for classical electronic receivers. 
Secondly, due to the abundant inherent energy levels that resonate with different radio frequencies (RF), Rydberg atoms can respond to a wide range of RF signals, ranging from several MHz~\cite{yang2024pcb} to GHz~\cite{zhang2024ultra} and even THz~\cite{lin2023terahertz}, which is impossible for classical electronic receivers. 
Thirdly, the physical length of the Rydberg atomic receiver is mainly determined by the length of the atomic vapor cell, which is independent of the wavelength of the received signal. This breaks the long-standing Chu's limit~\cite{kong1975theory} in antenna theory, and enables the fabrication of miniaturized atomic antenna arrays. 
Finally, since atomic quantum states are usually read out by optical approaches, electromagnetic crosstalk between Rydberg atomic receivers can be further suppressed, leading to a reduced array mutual coupling~\cite{yuan2025electromagnetic}. 
In summary, the expected sensitivity enhancement, large frequency tuning range, miniaturized array size, and reduced array mutual coupling make the Rydberg atomic receiver a promising technological candidate that could replace classical electronic receivers for future wireless communications and sensing.

\subsection{Prior works}
Rydberg atomic receivers are based on a quantum-optical phenomenon called the Autler-Townes (AT) effect, which was discovered by S. H. Autler and C. H. Townes in 1955~\cite{autler1955stark}. 
The AT effect of Rydberg atoms refers to the splitting phenomenon of optical transmissive peaks induced by external RF electric fields (E-fields). 
Since the AT peak separation is proportional to the applied E-field strength, the E-field strengths can be recovered by measuring the optical transmission spectrum of the atomic vapor.  
Due to its E-field sensing capability, the Rydberg atomic vapor serves as a general purpose RF receiver that has been quickly applied to E-field metrology~\cite{sedlacek2012microwave,ZHANG20241515,gordon2010quantum}, wireless sensing~\cite{zhang2023quantum,schmidt24rydberg}, and wireless communications~\cite{anderson2020atomic}. 

For wireless communications, experimental research efforts have been developed in conjunction with theoretical understanding of the atomic receiving mechanism. 
The development history of atomic receivers strongly resembles that of classical RF receivers, where the non-coherent amplitude detection of crystal radio detectors~\cite{douglas1981communications} was finally replaced by coherent amplitude-phase detection of modern IQ receivers with significantly improved sensitivity, frequency selectivity, and phase resolution. 
For non-coherent atomic communications, the Rydberg atoms were initially employed to demodulate amplitude-modulated (AM) waves by the authors of~\cite{anderson2020atomic}. 
This AM receiver was based on the theory that treated Rydberg atoms as amplitude-domain sensors~\cite{schmidt24rydberg}.  
Despite the simplicity of this amplitude-domain explanation of the atomic functionality, it leads to the misunderstanding that the atomic receiver is only capable of amplitude detection without phase detection~\cite{cui2025rydberg}. This misunderstanding was then corrected by the coherent detection theory in the literature by the authors of~\cite{simons2019rydberg,jing2020atomic,gong2024rydberg}. In these works, instead of being treated as an RF amplitude detector, the Rydberg atomic receiver acted as an atomic mixer that down-converted the incident RF signal into an optical intermediate frequency (IF) signal with one or multiple RF local oscillator (LO) signals. 
This superheterodyne approach experimentally improved the sensitivity of atomic receivers by three orders of magnitude~\cite{jing2020atomic}, and was further extended to approach the standard quantum limit (SQL)~\cite{tu2024approaching,chen2025harnessing}. 

Alongside the superiority of high sensitivity, another attractive benefit of Rydberg atomic receivers is their multi-band receiving capability. This capability allows for simultaneous reception of RF signals from multiple widely separated bands with a single atomic vapor cell. 
This multi-band reception capability is different from receiving multiple frequencies within a single band~\cite{liu2022deep}, and is also different from carrier aggregation technology~\cite{lin2013modeling} that requires separate RF front-end circuits for each different band. 
The multi-band reception of Rydberg atomic receivers was first demonstrated by the authors of~\cite{holloway2020multiple}, where two music-modulated RF signals were simultaneously received by mixing rubidium and cesium atoms inside a vapor cell. To reduce atom species, the authors of~\cite{du2022realization} realized RF reception in both Ku and Ka bands with only cesium atoms by exploiting their $66S_{1/2}\leftrightarrow 66P_{1/2}$ and $66S_{1/2}\leftrightarrow 67P_{3/2}$ transitions. To further expand the reception frequency range, the authors of~\cite{zhang2024ultra} adopted the atomic superheterodyne architecture~\cite{jing2020atomic} and demonstrated concurrent dual-band reception with a center frequency ranging from $300\,{\rm MHz}$ to $25\,{\rm GHz}$. Similar to classical electronic receivers, the frequency tuning of multi-band atomic receiver is performed by altering the LO configurations (LO frequencies and amplitudes).

\subsection{Motivation} 
Despite their benefits in simultaneous multi-band reception, multi-band atomic receivers are subject to intermediate frequency interference (IFI), which is conceptually analogous to inter-symbol interference (ISI) in single-carrier systems and inter-carrier interference (ICI) in orthogonal frequency-division multiplexing (OFDM) systems. Although signals from different bands are separated in the RF domain, they will interfere with each other in the optical IF domain after the atomic down-conversion process. Thus, this superimposed IF signal will lead to IFI between different bands, which cannot be fully distinguished by a single atomic receiver. To mitigate the IFI problem, the authors of~\cite{cui2025rydberg} and~\cite{du2022realization} adopted the frequency division multiple access (FDMA) scheme, where different users are scheduled to occupy different IF bands for inter-band orthogonality. However, since the IF cutoff frequency of atomic receivers is limited to several MHz~\cite{zhu2025general,du2022realization}, allocating users to different bands will quickly deplete the available IF bandwidths. In summary, the IFI problem poses a fundamental problem on the multi-user transmission performance of multi-band atomic receivers.

\subsection{Our contributions}
To address the fundamental problem of IFI, in this paper, we propose a multi-input multi-output (MIMO) architecture for multi-band quantum receivers by exploiting the additional spatial multiplexing of MIMO receivers\footnote{Simulation codes will be provided to reproduce the results in this paper: \url{http://oa.ee.tsinghua.edu.cn/dailinglong/publications/publications.html}. }. The contributions of this paper are summarized as follows. 

\begin{itemize}
    \item To overcome the IFI problem of multi-band atomic quantum receivers, we propose a multi-band Rydberg atomic quantum MIMO receiver architecture (RAQ-MIMO) that employs multiple quantum receivers to distinguish frequency bands in the space domain. In addition to simultaneously receiving signals from far-separated RF bands, the RAQ-MIMO receiver assigns different spatial combiners to different users in each band, thus allowing to serve more users with the same frequency resources. 
    \item For the RAQ-MIMO architecture, we provide the dynamic signal model from our previous work~\cite{zhu2025general} in the form of {\it quantum transconductances}. This signal model incorporates the decay rates of the Rydberg states and the RF/laser detunings, which is readily extendible to the multi-band regime. In addition to the {\it quantum transconductance}-based signal model, we further adopt noise models that jointly consider the BBR noise, electronic thermal noise, laser relative intensity noise (RIN), image frequency noise, etc. This physically compliant signal model supports the design of RAQ-MIMO receivers. 
    \item With the accurate physics-based signal model, we formulate the spectral efficiency (SE) maximization problem for RAQ-MIMO systems. Based on this formulation, we propose the quantum weighted minimum mean square error (qWMMSE) algorithm to maximize the SE of RAQ-MIMO systems. Specifically, the non-convex SE maximization problem is first converted to a convex optimization problem with additional optimizable variables of LO configurations. Then, an alternate optimization framework is proposed to iteratively optimize the LO configurations, precoding matrices, and combining matrices.   
    \item Simulation results are provided to demonstrate the improved SE performance of the RAQ-MIMO system, with the proposed qWMMSE algorithm applied to both space division multiple access (SDMA) and frequency division multiple access (FDMA) schemes. In both multiple access schemes, the multi-band LO configurations are automatically balanced according to the channel qualities of each band and each user. Furthermore, numerical results have shown that RAQ-MIMO systems could possibly outperform classical electronic receiver-based MIMO systems by avoiding the mutual coupling effect of classical antennas. 
\end{itemize}

\subsection{Organization and Notation}
\emph{Organization}:
Section~\ref{sec2} introduces the system model of multi-band Rydberg atomic receivers, and formulates the SE maximization problem. Section~\ref{sec3} presents the problem transformation process to a convex optimization problem, which is then solved by the proposed qWMMSE algorithm. Section~\ref{sec4} presents the SE optimization performance. Finally, conclusions are drawn in Section~\ref{sec5}. 

\emph{Notation}: 
Bold uppercase characters ${\bf X}$ denote matrices, with $[{\bf X}]_{mn}$ representing its $(m,n)$--th entry; 
bold lowercase characters ${\bf x}$ denote vectors;
${\bf I}_n$ denotes the identity matrix of size $n$; 
${\bf X}\H, {\bf X}\T$, and ${\bf X}^*$ denotes Hermitian transpose, transpose, and complex conjugate of ${\bf X}$, respectively; 
for two operators $\mathcal{A}_1$ and $\mathcal{A}_2$, $[\mathcal{A}_1, \mathcal{A}_2]$ denotes the commutator $\mathcal{A}_1\mathcal{A}_2 - \mathcal{A}_2\mathcal{A}_1$, and $\{\mathcal{A}_1, \mathcal{A}_2\}$ denotes their anti-commutator $\mathcal{A}_1\mathcal{A}_2+\mathcal{A}_2\mathcal{A}_1$;  
${\bf M}_{ij}$ denotes the elementary matrix with the only ``1'' located at the $(i,j)$-th entry; 
$\otimes$ denotes the Kronecker product of two matrices;  
$\rm{i}$ is the imaginary unit; 
$\hbar$ denotes the reduced Planck's constant; 
$c_0$ denotes the speed of light in a vacuum; 
$\epsilon_0$ denotes the vacuum permittivity; 
$\eta_0$ denotes the vacuum wave impedance; 
${\rm vec}({\bf X})$ stacks the columns of the matrix $\bf X$ into a single column vector, and ${\rm unvec}({\bf x})$ undoes this operation; 
${\rm PSD}[X(t)]$ denotes the double-sided power spectral density of the random process $X(t)$; 
$\delta_{mn}$ is the Kronecker delta that evaluates to $1$ only when $m=n$; 
$\mathcal{U}(a,b)$ denotes the uniform distribution from $a$ to $b$.

\section{System Model} \label{sec2}
In this section, we introduce the system model of the multi-band Rydberg atomic receiver, and formulate the RAQ-MIMO SE maximization problem.  
\begin{figure}[t]
    \centering
    \includegraphics[width=\figsize\linewidth]{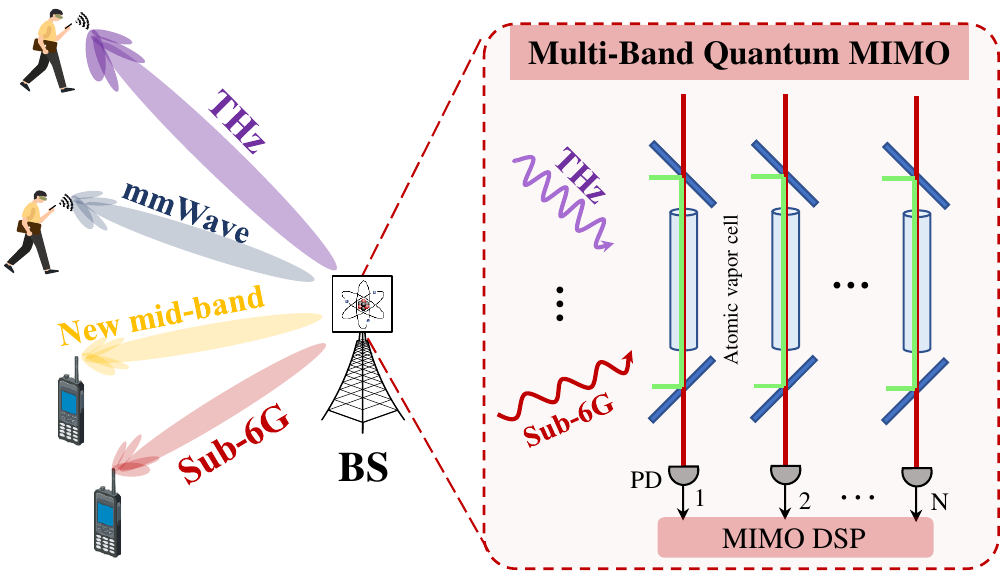}
    \caption{Uplink RAQ-MIMO communications enabled by Rydberg atomic receiver array. }
    \label{fig:Conceptual_MultiBandReceiver}
\end{figure}

\subsection{Overview of existing signal and noise models} 
Communication engineers need physically accurate signal models for future elaborated design in information transmission. Different from classical electronic receivers that are already familiar to the communications community, Rydberg atomic receivers are relatively not well-known to most of the communication engineers. Following the construction principles of the already mature classical signal models, here we propose several criteria for a physically correct, mathematically reasonable, and practically applicable model for quantum receivers. These criteria are also applicable to more advanced receiver architectures that will appear in the future. 
\begin{itemize}
    \item {\bf Correctness in dimension}. Every physical quantity has its dimension, and can be reduced to a combination of seven basic physical quantities (international system of units, SI). Ensuring correct dimension of each physical quantity will help avoid magnitude errors that are easy to encounter in numerical simulations. For example, all digital signals are dimensionless, the E-field strengths are in ${\rm V/m}$, and the input/output impedances of electronic components are in $\rm \Omega$. 
    \item {\bf Frequency range}. Each signal model has a frequency range in which it is valid or approximately valid. For a communication system, the frequency range is usually jointly determined by the operational bands of the antennas, RF amps/filters, IF amps/filters, and the ADCs. 
    \item {\bf Amplitude range}. Although most physical systems are linear in the small-signal regime~\cite{dal2019linear}, they usually exhibit non-linear effects when the input signal amplitude is relatively large. Non-linear effects will degrade the output signals or interfere with neighboring devices by creating new frequencies. For these reasons, the non-linear effects should be considered in applications with large dynamic range of signal amplitude. 
\end{itemize}
Compared with these criteria, the existing models for Rydberg atomic receivers are summarized in {\bf Table~\ref{tab:Signal_models}}. In this paper, to ensure physical compliance, we adopt our quantum transconductance model~\cite{zhu2025general} in the following design and optimization of RAQ-MIMO systems. 

\begin{table*}[t]
    \centering
    \begin{threeparttable} 
        \caption{Comparison of different signal models for Rydberg atomic receivers} \label{tab:Signal_models}
        \vspace{-3pt}
        \renewcommand{\arraystretch}{1.5}
        \begin{tabular}{|c|c|c|c|}
            \hline 
            {\bf Model criteria}             & Probe-readout model~\cite{gong2024rydberg,gong2025equivalent,chen2025harnessing}  & Rabi-readout model~\cite{cui2025towards}   &  \makecell{Quantum transconductance \\model~\cite{zhu2025general}}  \\ 
            \hline
            Physical dimensions & Mostly ensured &  Ensured  &  Clearly stated      \\ 
            \hline
            Frequency range     & \makecell{RF tunable \\ range discussed} & \makecell{RF tunable \\ range discussed} & \makecell{IF dynamic \\  range discussed } \\  
            \hline
            Amplitude range     & Large/small signal & Large/small signal & Mainly small-signal \\ 
            \hline
            \makecell{Consistency with \\experimental results} & Not discussed  & Not discussed & \makecell{$\kappa$ factor \\ consistent} \\
            \hline
            \makecell{Signal readout\\mechanism} & \makecell{Direct PD \\conversion} & \makecell{AT splitting\\ spectral measurement} & \makecell{Direct PD \\conversion} \\ 
            \hline 
            \makecell{Time-domain \\property}      & Static & Static & Dynamic  \\ 
            \hline 
            \makecell{Frequency-domain \\ property} & N/A & N/A & $g_q(\ri \omega)$ \\ 
            \hline 
            \bf{\makecell{Main source \\ of inaccuracy}} & In-band BBR noise & Spectral measurement & \makecell{Multi-atom interaction, \\ Thermal Doppler effect} \\ 
            \hline 
        \end{tabular}
    \end{threeparttable}
\end{table*}

\subsection{Basic principles for quantum receivers}
Let us consider a Rydberg atomic receiver with the atomic vapor cell exposed to $M$ different RF bands simultaneously, as is shown in Fig.~\ref{fig:Conceptual_MultiBandReceiver}. The center frequency and bandwidth of the RF signal in the $m$-th band is respectively denoted as $f_{c,m}$ and ${\rm BW}_{m}$, where $1\leq m \leq M$. To enable both amplitude and phase detection of RF signals, a local oscillator (LO) signal of frequency $f_{c,m}$ and electric field intensity $E_{{\rm LO},m}\,{\rm [V/m]}$ is applied to each of the $M$ bands, together with the incident information-carrying signal $E_{{\rm sig},m}\,{\rm [V/m]}$.  

Generally speaking, the $M$-band atomic receiver translates the change of $E_{{\rm sig},m}$ into its varying probe light transmission coefficient $T_p=T_p(t)$, where $t$ denotes time. After photo-electric conversion by the photodetector, this change in $T_p$ is reflected in the output photocurrent of the photodetector, which enables subsequent electronic domain signal processing. This E-field-to-photocurrent transfer function was studied in our previous work~\cite{zhu2025general}, which is denoted as quantum transconductances $\{g_{q,m}\}_{m=1}^M\,{\rm [\Omega]^{-1}}$. The value of $g_{q,m}$ reflects the level of sensitivity of the alkali atomic vapor to the change of external signal field $E_{{\rm sig},m}, \forall m$. Specifically, the analytic representation\footnote{Also known as the complex baseband representation of a real-valued bandpass signal.} of the output photocurrent signal $\Delta I_{\rm ph}(t)$ of the RAQR is linearly determined by the input E-fields of each band to be   
\begin{equation}
    \Delta I_{\rm ph}(t) = L \sum_{m=1}^M g_{q,m} E_{{\rm sig},m}(t), 
\end{equation}
where $L{\rm \,[m]}$ is the length of the atomic vapor cell, and $E_{{\rm sig},m}(t)\,{\rm [V/m]}$ is the analytic representation of the received RF E-field in the $m$-th band. 

\begin{figure}
    \centering
    \includegraphics[width=\figsize\linewidth]{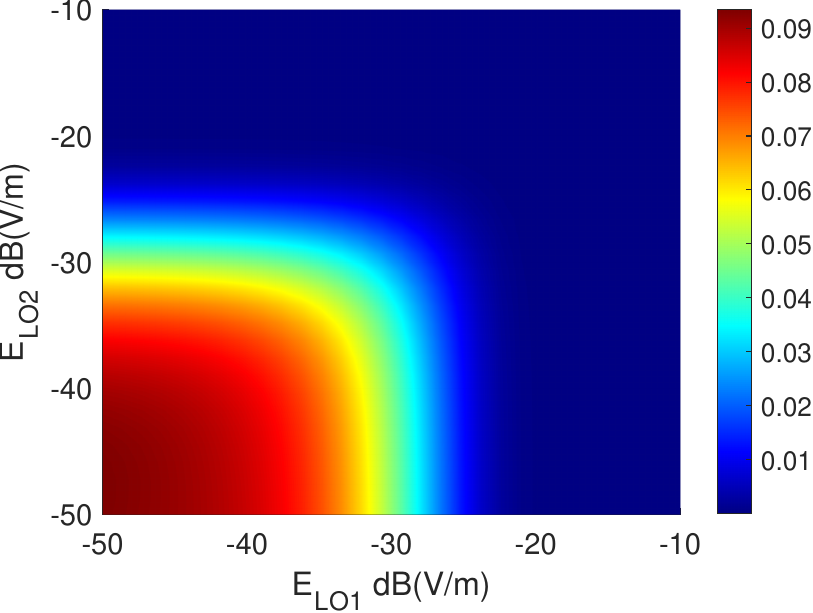}
    \caption{Probe light transmission coefficient $T_p$ of the atomic vapor v.s. LO E-fields $E_{{\rm LO},1}$ and $E_{{\rm LO},1}$ in a dual-band ($M=2$) Rydberg atomic receiver system. The values of $T_p$ are computed with the zero-detuning assumption on all the EM fields. }
    \label{fig:Technical_PT_wrt_ELOs}
\end{figure}

\begin{figure}[t]
    \centering
    \includegraphics[width=\figsize\linewidth]{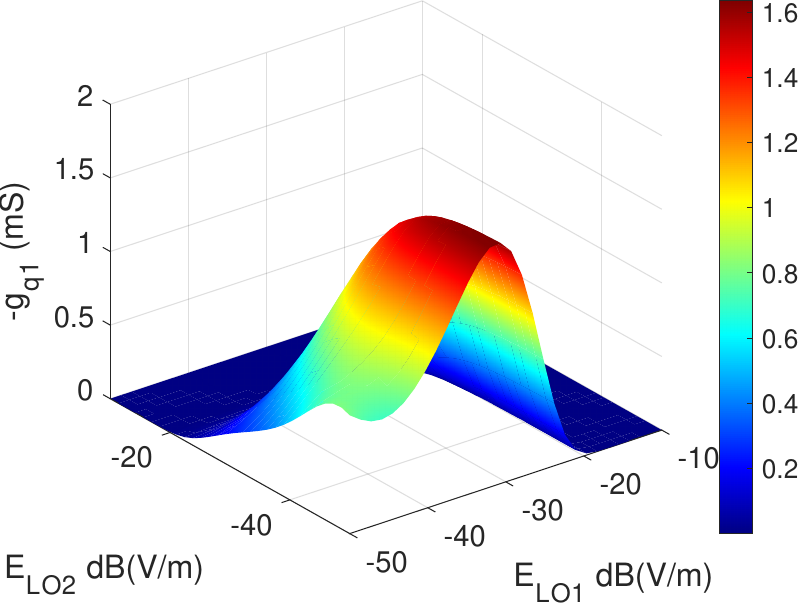}
    \caption{Quantum transconductance $g_{q,1}$ in band 1 as a function of LO E-field intensities in a dual-band ($M=2$) Rydberg atomic receiver system. }
    \label{fig:Technical_gq1_wrt_ELO}
\end{figure}

\begin{figure}[t]
    \centering
    \includegraphics[width=\figsize\linewidth]{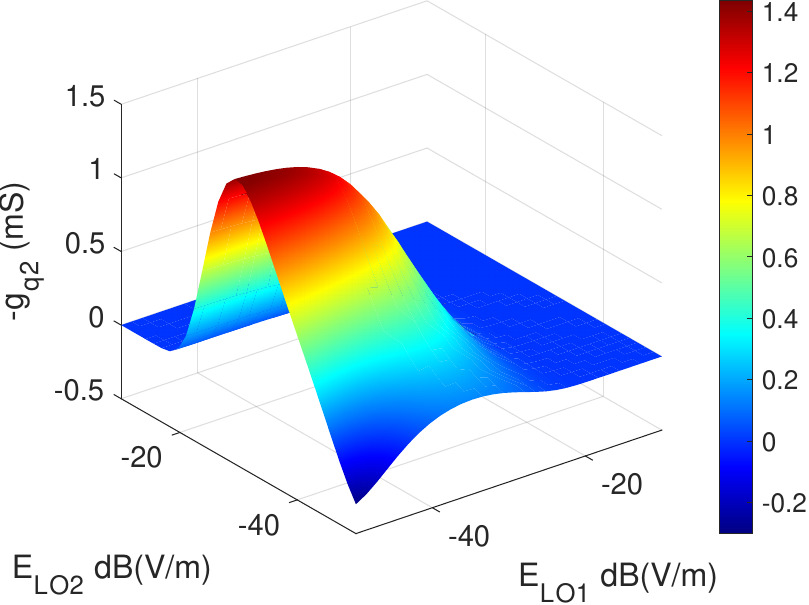}
    \caption{Quantum transconductance $g_{q,2}$ in band 2 as a function of LO E-field intensities in a dual-band ($M=2$) Rydberg atomic receiver system. }
    \label{fig:Technical_gq2_wrt_ELO}
\end{figure}

Fig.~\ref{fig:Technical_PT_wrt_ELOs} shows the dependence of the probe light transmission coefficient $T_p$ on the LO E-field intensities $E_{{\rm LO},m}$. It can be observed that the transmission $T_p$ drops as external fields increases, showcasing the quantum EIT-AT effect~\cite{autler1955stark} in the dual-LO regime. By taking the partial derivatives of Fig.~\ref{fig:Technical_PT_wrt_ELOs} on the operating point $\{E_{{\rm LO},m}\}_{m=1}^M$ and applying some appropriate coefficients, the quantum transconductances $g_{q,m}$ are evaluated and shown in Fig.~\ref{fig:Technical_gq1_wrt_ELO} and Fig.~\ref{fig:Technical_gq2_wrt_ELO}. It can be observed that the peak values of $g_{q,1}$ and $g_{q,2}$ for this dual-band quantum receiver do not appear at the same LO operating point $(E_{{\rm LO},1}, E_{{\rm LO},2})$~\cite{you2023exclusive}. For example, the operating point that maximizes $g_{q,1}$ in the first band may result in a small $g_{q,2}$, thus degrading the communication performance in the second band.  
Therefore, an optimization algorithm is needed to balance the gains in each band to achieve an optimal overall communication performance. 

\subsection{Detailed principles for multi-band quantum receivers}
In this subsection, we introduce how to compute the quantum transconductances $\{g_{q,m}\}_{m=1}^M$ from the physical parameters of the atomic vapor. 

The incidence of the LO field $E_{{\rm LO},m}$ to the atomic vapor drives the Rydberg-Rydberg electron transition $\ket{3}\leftrightarrow \ket{m+3}$, where $\ket{1}, \ket{2}, \ket{3}$ denotes the ground state, the intermediate excited state, and the first in-operation Rydberg state of the alkali atoms. The subsequent Rydberg states are denoted as $\ket{m+3},1\leq m\leq M$. 

The evolution of the atomic quantum state is mainly determined by the interaction Hamiltonian. Assume the transition dipole moment of $\ket{3}\leftrightarrow \ket{m+3}$ is $\mu_{{\rm RF},m}$, then the transition Rabi frequency is $\Omega_{{\rm RF},m} = \mu_{{\rm RF},m} E_{{\rm RF},m}/\hbar$. The interaction Hamiltonian ${\bf H}\in\mathbb{C}^{(M+3)\times(M+3)}$ is expressed in~\eqref{eqn:Hamiltonian}, which is jointly determined by the RF wave parameters $\{(\Omega_{{\rm RF},m}, \Delta_{{\rm RF},m})\}_{m=1}^M$, the probe light parameters $(\Omega_p,\Delta_p)$, and the control light parameters $(\Omega_c, \Delta_c)$. 
\begin{figure*}
\begin{equation}
    {\bf H} = \begin{bmatrix} 0 & \Omega_p/2 & 0 & 0 & 0 & \cdots & 0 \\
    \Omega_p/2 & -\Delta_p & \Omega_c/2 & 0 & 0 & \cdots & 0 \\
    0 & \Omega_c/2 & -\Delta_p-\Delta_c & \Omega_{{\rm RF},1}^*/2 & \Omega_{{\rm RF},2}^*/2 & \cdots & \Omega_{{\rm RF},M}^*/2 \\ 
    0 & 0 & \Omega_{{\rm RF},1}/2 & -\Delta_p-\Delta_c-\Delta_{{\rm RF},1} & 0 & \cdots & 0 \\ 
    0 & 0 & \Omega_{{\rm RF},2}/2 & 0 & -\Delta_p-\Delta_c-\sum_{m=1}^{2}\Delta_{{\rm RF},m} &\cdots & 0 \\
    \vdots & \vdots & \vdots & \vdots & \vdots & \ddots & \vdots \\
    0 & 0 & \Omega_{{\rm RF},M}/2 & 0 & 0 & \cdots & -\Delta_p-\Delta_c-\sum_{m=1}^M \Delta_{{\rm RF},m}
    \end{bmatrix} \label{eqn:Hamiltonian}
\end{equation}
\end{figure*}

The quantum decay and decoherence can be described by the decay parameters. Specifically, let $\gamma_n$ be the decay rate of the quantum state $\ket{n}, 2\leq n\leq M+3$. The decay rate matrix $\bf \Gamma$ is defined as ${\bf \Gamma} = {\rm diag}(0, \gamma_2, \gamma_3, \cdots, \gamma_{M+3})$. Then, the Lindblad operator is written as 
\begin{equation}
\begin{aligned}
    \mathcal{L}[{\bm \rho}] &= -\frac{1}{2}\{{\bf \Gamma}, {\bm \rho}\} + \left(\gamma_2\rho_{22}+\sum_{m=1}^M\gamma_{m+3}\rho_{m+3,m+3}\right){\bf M}_{11} \\
    &~~~~+ \gamma_3\rho_{33}{\bf M}_{22}, 
    \end{aligned}
\end{equation}
which determines the time evolution of the density matrix ${\bm \rho}$ via the master equation $\diff{\bm \rho}/\diff t = -\ri [{\bf H},{\bm \rho}] + \mathcal{L}[{\bm \rho}]$. 

{\bf Steady-state response of the quantum system.} In the case where the complex envelope of the applied external RF signal $E_{{\rm sig},m}(t)$ is slow-varying compared to the Rabi frequencies $\{\Omega_{{\rm LO},m} , \Omega_p, \Omega_c\}$, the steady-state approximation $\diff{\bm\rho}/\diff t={\bf 0}$ can be applied, rendering the steady-state density matrix $\bar{\bm\rho}$ to satisfy a homogeneous linear equation~\cite{zhu2025general} 
\begin{equation}
    {\bf A}_0 \bar{\bf x} = {\bf 0},
\end{equation}
where $\bar{\bf x}={\rm vec}(\bar{\bm \rho})\in\mathbb{C}^{(M+3)^2\times 1}$ is the vectorized steady-state density matrix, and ${\bf A}_0\in\mathbb{C}^{(M+3)^2\times (M+3)^2}$ is linearly related to ${\bf H}$ and $\bf \Gamma$ via 
\begin{equation}
\begin{aligned}
    {\bf A}_0 =& -\ri({\bf I}_{M+3}\otimes{\bf H}_0 - {\bf H}_0\T\otimes {\bf I}_{M+3}) \\ 
    &~-\frac{1}{2}({\bf \Gamma}\otimes{\bf I}_{M+3}+{\bf I}_{M+3}\otimes{\bf \Gamma})\\
    &~+ \gamma_2{\bf M}_{1,(M+3)+2}+\gamma_3{\bf M}_{(M+3)+2,2(M+3)+3}\\
    &~+\sum_{m=1}^M\gamma_{m+3}{\bf M}_{1,(m+2)(M+3)+(m+3)},
\end{aligned}\label{eqn:A0}
\end{equation}
where ${\bf H}_0$ is the zero-input version of the Hamiltonian matrix in ${\bf H}$~\eqref{eqn:Hamiltonian} with all the signal E-fields set to zero. Note that although the decay rates $\{\gamma_n\}_{n=2}^{M+3}$ are usually three orders of magnitude smaller than the probe/control Rabi frequencies $\Omega_{p,c}$ and $\{\Omega_{{\rm RF},m}\}_{m=1}^M$, we discover in our previous work~\cite{zhu2025general} that these decay rates significantly affect the probe response determined by $\Im{{\bm \rho}_{21}}$. Thus, we incorporate the influence of these decay rates in the expression of ${\bf A}_0$. 

{\bf Dynamic response of the quantum system.} Following our previous work on the general signal model for Rydberg atomic receivers~\cite{zhu2025general}, the atomic response to the RF signal fields can be fully characterized by a novel physical quantity called the {\it quantum transconductance}\footnote{Note that this quantum transconductance $g_{q,m}$ here is defined in the low-IF regime. The complete atomic frequency response $g_{q,m}(\ri \omega)$ that spans the entire IF bandwidth is studied in our previous work~\cite{zhu2025general}.} $\{g_{q,m}\}_{m=1}^M$, which is given by 
\begin{equation}
    g_{q,m} = \bar{I}_{\rm ph} \cdot \frac{2k_p N_0\mu_{12}^2}{\epsilon_0\hbar\Omega_p}\frac{\partial\Im{[\bar{\bm \rho}]_{21}}}{\partial {E}_{{\rm LO},m}} \quad{\rm [\Omega^{-1}]},\quad 1\leq m\leq M, \label{eqn:q_tcond}
\end{equation}
where $\bar{I}_{\rm ph}$ is the zero-input photocurrent when no signal is applied, $k_p=2\pi/\lambda_p\,{\rm [m^{-1}]}$ is the wavenumber of the probe light, $N_0\,{\rm [m^{-3}]}$ is the atomic density inside the vapor cell, and $\mu_{12}\,{\rm [C\cdot m]}$ is the transition dipole between the ground state $\ket{1}$ and the intermediate excited state $\ket{2}$. Note that quantum transconductance $g_{q,m}$ has a dimension of conductance in Siemens $\rm [S]=[\Omega]^{-1}$.

Using $\Im{\bar{\bm \rho}_{21}}=[\bar{\bf x}]_2$ and the steady-state equation ${\bf A}_0\bar{\bf x}={\bf 0}$, we can evaluate the partial derivative ${\partial\Im{[\bar{\bm \rho}]_{21}}}/{\partial {E}_{{\rm LO},m}}$ explicitly. Following the same approach in~\cite{zhu2025general}, we first convert the singular equation ${\bf A}_0\bar{\bf x}={\bf 0}$ into the non-singular form ${\bf C}_0\bar{\bf z} + {\bf w}_0 /\sqrt{M+3}={\bf 0}$ by exploiting the trace-preserving property of ${\bf A}_0$, leading to a reduced matrix size from $(M+3)^2$ to $(M+3)^2-1$. The matrices ${\bf C}_0, {\bf w}_0$ are determined by 
\begin{equation}
    {\bf Q}\T{\bf A}_0{\bf Q} = \begin{bmatrix}
        0 & {\bf 0}_{1\times ((M+3)^2-1)} \\ 
        {\bf w}_0 & {\bf C}_0
    \end{bmatrix},
\end{equation}
where ${\bf Q} = [{\bf u}_{M+3}, {\bf q}_1,\cdots {\bf q}_{(M+3)^2-1}]\in\mathbb{R}^{(M+3)^2\times (M+3)^2}$ is an orthogonal matrix with the first column ${\bf u}_{M+3}$ set to be $[{\bf u}_{M+3}]_i=1/\sqrt{M+3},\,i=(M+3)(j-1)+j,j=1,2,\cdots,M+3$, and the rest columns arbitrarily chosen to ensure ${\bf Q}\T{\bf Q}={\bf I}_{(M+3)^2}$. After solving the non-singular equation ${\bf C}_0\bar{\bf z} + {\bf w}_0 /\sqrt{M+3}={\bf 0}$ for $\bar{\bf{z}}$, the value of $[\bar{\bm\rho}]_{21}$ can be recovered as 
\begin{equation}
    [\bar{\bm \rho}]_{21} = \left[{\bf Q}\begin{bmatrix} \frac{1}{\sqrt{M+3}} \\  \bar{\bf z}\end{bmatrix}\right]_{2}. \label{eqn:rho_as_z}
\end{equation}
The analysis pipeline from the LO amplitudes to the quantum transconductances is illustrated by 
\begin{equation}
\begin{aligned}
    \{E_{{\rm LO},m}\}_{m=1}^M &\overset{\eqref{eqn:Hamiltonian},\eqref{eqn:A0}}{\Longrightarrow} {\bf A}_0 \overset{{\bf Q}}{\Longrightarrow}({\bf C}_0, {\bf w}_0) \\
    & \overset{\eqref{eqn:rho_as_z}}{\Longrightarrow} \left([\bar{\bm \rho}]_{21},\,\frac{\partial\Im{[\bar{\bm \rho}]_{21}}}{\partial E_{{\rm LO},m}} \right)\\
    &\overset{\eqref{eqn:q_tcond}}{\Longrightarrow} \{g_{q,m}\}_{m=1}^M. 
\end{aligned}
\end{equation}

{\bf Baseband equivalent model of the quantum system.} After obtaining the quantum transconductance $g_{q,m}$, the single-input single-output (SISO) signal model for each band $m$ is written as 
\begin{equation}
    y_m(t) = \sqrt{\frac{P_{\rm T}}{P_{{\rm qref},m}}} Hx_{{\rm BB},m}(t) + w_{{\rm BB},m}(t),
    \label{eqn:signal_model_single_band}
\end{equation}
where 
\begin{equation}
\begin{aligned}
    \frac{1}{\sqrt{P_{{\rm qref},m}}} &= \underbrace{\frac{1}{2V_{\rm ref}}}_{\rm ADC}\cdot \underbrace{R_{\rm T}K_c}_{\rm TIA}\cdot \underbrace{L g_{q,m}}_{\rm atomic}\times \underbrace{\sqrt{\frac{8\pi \eta_0}{\lambda_{c,m}^2}}}_{\text{dimension conversion}} \\
    &:= g_{q,m}C_{{\rm sig},m}
\end{aligned} \label{eqn:quantum_reference_power}
\end{equation}
is the quantum reference power coefficient~\cite{zhu2025general} that is proportional to the quantum transconductances of the Rydberg atomic receiver, $x_{{\rm BB},m}(t)$ is the unit-power transmitted baseband signal, $P_{\rm T}\,{\rm [W]}$ is the transmitted power of the classical transmitter, and $w_{{\rm BB},m}(t)$ is the complex baseband noise process with power spectral density determined by~\cite[Eq.~(61-64)]{zhu2025general}. The noise process $w_{{\rm BB},m}(t)$ is the total effect of various noise sources, where the noise sources of the quantum receiver are listed in {\bf Table~\ref{tab:Noise_Sources}}. Note that the definition of the channel coefficient $H$ in the quantum signal model~\eqref{eqn:signal_model_single_band} is the same as those of classical signal models. The symbols $V_{\rm ref}$, $R_{\rm T}$ and $K_c$ are the ADC reference voltage, the transimpedance of the transimpedance amplifier (TIA) which amplifies the photocurrent, and the current dividing coefficient of the electronic stages that processes the current signal generated by the photodiode~\cite{zhu2025general}, respectively. The symbol $\lambda_{c,m}$ denotes the carrier wavelength of the $m$-th band.   

\begin{table*}[t]
    \centering
    \begin{threeparttable} 
        \caption{Noise sources in the quantum receiver} \label{tab:Noise_Sources}
        \vspace{-3pt}
        \setstretch{1.07}
        \begin{tabular}{|l|c|c|c|c|}
            \hline 
            {\bf Noise sources}         & {\bf Influence}   & Our works     & \cite{gong2024rydberg,gong2025equivalent}&\cite{cui2025towards,cui2025rydberg}  \\ 
            \hline
            In-band BBR                 & Strong            & \checkmark    &            & \checkmark\\ 
            \hline
            Image frequency BBR        &Strong (3dB$^\star$)& \checkmark    &            &  \\ 
            \hline 
            Electronic amplifier noise  & Strong            & \checkmark    & \checkmark & \checkmark\\ 
            \hline
            Laser relative intensity noise & Moderate       & \checkmark    &            & \\ 
            \hline
            Out-of-band BBR             & Weak$^\dagger$    & \checkmark    & \checkmark & \\
            \hline 
            Photon shot noise           & Laser-dependent   & \checkmark    & \checkmark & \checkmark \\
            \hline 
            Quantum projection noise    & Very weak         &               & \checkmark & \checkmark \\ 
            \hline 
        \end{tabular}
        {\footnotesize $^\star$: The BBR noise in the image sideband will enter the receiver, inreasing the total BBR noise level by a factor of 3\,dB; $\dagger$: The out-of-band BBR hardly contribute to the received in-band noise, however, it significantly reduces the signal gain of the quantum stage. }
    \end{threeparttable}
\end{table*}

\subsection{Multi-band uplink MU-MIMO signal model}

Consider a multi-band uplink multi-user MIMO system with Rydberg atomic receivers at BS. Specifically, we assume that the BS is equipped with a Rydberg atomic receiver array with $N_r$ vapor cells. Each of the vapor cells operates in $M$ disjoint bands. In each band there are $K$ users, where each user is equipped with $N_t$ transmit antennas. For each user, a number of $S=\min\{N_t, N_r\}$ data streams are transmitted. 

\begin{figure}[t]
    \centering
    \includegraphics[width=\figsize\linewidth]{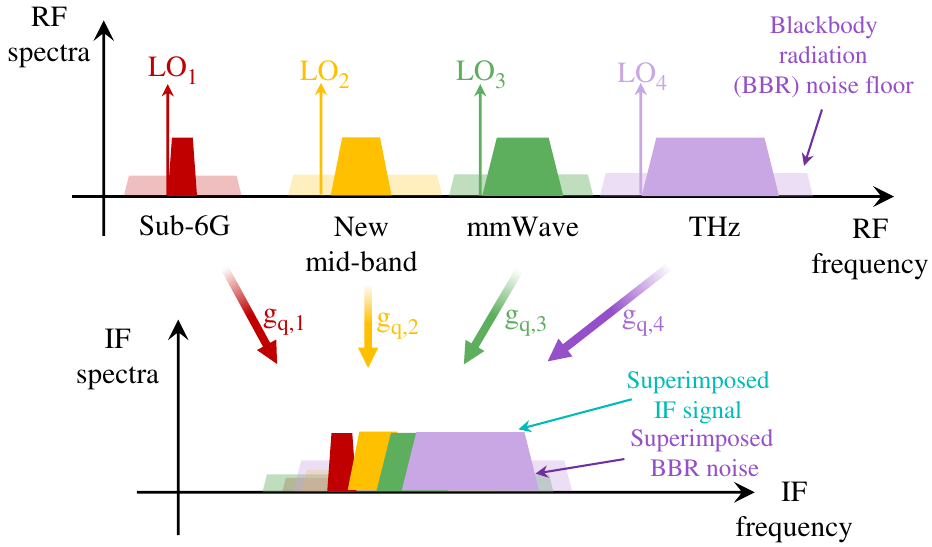}
    \caption{RF to IF down-conversion process of multi-band Rydberg atomic receiver. The IFI phenomenon is clearly shown. }
    \label{fig:Conceptual_RF2IF}
\end{figure}

{\bf Signal components.} Although users in different bands transmit RF signals at different carrier frequencies $\{f_{c,m}\}_{m=1}^M$, after down-conversion by their own LO tone $\{E_{{\rm LO},m}\}_{m=1}^M$ to the IF, the IF signals from different RF bands will be superimposed together in the atomic receiver, as shown in Fig.~\ref{fig:Conceptual_RF2IF}. Since the signals from different RF bands cannot be separated in the IF domain, they can only be distinguished in the spatial domain. In this spatial division multiple access (SDMA) system, the uplink transmission model is given by 
\begin{equation}
    {\bf y}^{\rm SDMA} = \sum_{m=1}^{M} \frac{1}{\sqrt{P_{{\rm qref},m}}}\sum_{k=1}^K {\bf H}_{m,k}{\bf V}_{m,k} {\bf s}_{m,k} + {\bf w}_{\rm tot},
    \label{eqn:MU-MIMO_signal_model}
\end{equation}
where ${\bf s}_{m,k}\in \mathbb{C}^{S\times 1}$ is the transmit symbol vector of the $k$-th user in the $m$-th band with unit-variance entries, ${\bf V}_{m,k}\in\mathbb{C}^{N_t\times S}$ is the precoding matrix that satisfies the power constraint $\tr({\bf V}_{m,k}{\bf V}_{m,k}\H)\leq P_{m,k} \,{\rm [W]}$, ${\bf H}_{m,k}\in\mathbb{C}^{N_r\times N_t}$ is the uplink channel matrix, $P_{{\rm qref},m}\,{\rm [W]}$ is the quantum reference power of the $m$-th band, ${\bf w}_{\rm tot}\in\mathbb{C}^{N_r\times 1}$ is the total noise vector of the quantum receiver, and ${\bf y}^{\rm SDMA}\in\mathbb{C}^{N_r\times 1}$ is the signal vector received by the BS. 

{\bf Noise components.} For the noise vector, its distribution is modeled by ${\bf w}_{\rm tot}\sim \mathcal{CN}({\bf 0}, {\bf C}_{\rm tot})$, and the total noise covariance matrix can be decomposed as 
\begin{equation}
    {\bf C}_{\rm tot} = \sigma_e^2 {\bf I} + \sum_{m=1}^M (g_{q,m}C_{{\rm sig},m})^2 {\bf C}_{q,m},
    \label{eqn:quantum_noise_components}
\end{equation}
where $\sigma_e^2$ is the variance of the internal electronic noise of the quantum receiver (dimensionless constant), $g_{q,m}\,{\rm [\Omega^{-1}]}$ is the quantum transconductance of the $m$-th band that depends on the quantum LO bias, $C_{{\rm sig},m}\,{\rm [\Omega/\sqrt{W}]}$ is the signal dimension conversion coefficient of the $m$-th band that depends on the center frequency $f_{c,m}$ of each RF band, and ${\bf C}_{q,m}\,{\rm [W]}$ is the BBR noise covariance matrix given by 
\begin{equation}
    {\bf C}_{q,m} = \frac{4}{3}k_{\rm B}T\cdot {\rm BW}_m \cdot \zeta(\ell)\hat{\bf C}_{q,m}, 
\end{equation}
where $\hat{\bf C}_{q,m}$ is the normalized BBR correlation matrix with 1's on its diagonal, ${\rm BW}_m$ is the bandwidth of the $m$-th RF band, and $\zeta(\ell), \ell=L/\lambda_{c,m}$ is the BBR coherence factor~\cite{zhu2025general} that quantifies the spatial coherence of the BBR E-field. This formula is proved in {\bf Appendix~\ref{app:BBR_cov_formula}}. Note that ${\rm BW}_m\leq {\rm BW}_{\rm IF}, \forall m$ should be satisfied for effective signal reception, where ${\rm BW}_{\rm IF}$ denotes the optical IF bandwidth of the Rydberg atomic receiver~\cite{zhu2025general}. 

We note the following facts about the uplink MU-MIMO SDMA model~\eqref{eqn:MU-MIMO_signal_model} and the noise model~\eqref{eqn:quantum_noise_components}. 
\begin{enumerate}
    \item The sum expression of the BBR noises in~\eqref{eqn:quantum_noise_components} is due to the superposition of the down-converted BBR noises in each RF band to the same IF band; 
    \item The quantum transconductance of a Rydberg atomic vapor cell is physically analogous to the transconductance of a microwave transistor; 
    \item By adopting the signal model formulation~\eqref{eqn:MU-MIMO_signal_model}, the entries of the channel matrix ${\bf H}_{m,k}$ is identical to the original channel coefficients for classical electronic receivers, making the signal and noise models fully compatible with existing classical models. 
\end{enumerate}

\subsection{Formulation of the weighted SE maximization problem} 
By using the relationship $1/\sqrt{P_{{\rm qref},m}} = g_{q,m}C_{{\rm sig},m}$ and decomposing the total noise into each band, the SDMA signal model~\eqref{eqn:MU-MIMO_signal_model} is further structured as 
\begin{equation}
\begin{aligned}
    {\bf y}^{\rm SDMA} &= \sum_{m=1}^M g_{q,m}C_{{\rm sig},m}\left(\sum_{k=1}^K {\bf H}_{m,k}{\bf V}_{m,k}{\bf s}_{m,k} + {\bf w}_{q,m}\right) \\
    &+ {\bf w}_e,
\end{aligned}
\end{equation}
where ${\bf w}_{q,m}\sim\mathcal{CN}({\bf 0}, {\bf C}_{q,m})$, $\forall m$ is the BBR noise vector, and ${\bf w}_e\sim\mathcal{CN}({\bf 0}, \sigma_e^2{\bf I})$ is the internal electronic noise vector. We further assume that the combiner matrix for the $k$-th user in the $m$-th band is ${\bf U}_{m,k}\H\in \mathbb{C}^{S\times N_r}$. Then the MIMO signal transfer model for the $(m,k)$-th user is written as 
\begin{equation}
\begin{aligned}
    \hat{\bf s}_{m,k}^{\rm SDMA} & = {\bf U}_{m,k}\H {\bf y}^{\rm SDMA} \\ 
    &= \underbrace{g_{q,m}C_{{\rm sig},m}{\bf U}_{m,k}\H {\bf H}_{m,k}{\bf V}_{m,k}{\bf s}_{m,k}}_{\text{desired signal}} \\
    &~+ \underbrace{\sum_{(n,\ell)\neq (m,k)}g_{q,n}C_{{\rm sig},n}{\bf U}_{m,k}\H{\bf H}_{n,\ell}{\bf V}_{n,\ell}{\bf s}_{n,\ell}}_{\text{interference from other bands/users}} \\
    &~+ \underbrace{{\bf U}_{m,k}\H\left(\sum_{n=1}^M g_{q,n}C_{{\rm sig},n}{\bf w}_{q,n} + {\bf w}_e\right).}_{\text{noise}}
\end{aligned} \label{eqn:Linear-Detector}
\end{equation}
The interference-plus-noise covariance matrix ${\bf R}_{m,k}^{\rm SDMA}$ of the received vector for the $(m,k)$-th user is given by 
\begin{equation}
    {\bf R}_{m,k}^{\rm SDMA} = {\bf R}_{\bf y}^{\rm SDMA} - (g_{q,m}C_{{\rm sig},m})^2 {\bf H}_{m,k}{\bf V}_{m,k}{\bf V}_{m,k}\H{\bf H}_{m,k}\H,
\end{equation}
where the total covariance matrix of the received signal ${\bf R}_{\bf y}$ is given by 
\begin{equation}
\begin{aligned}
    {\bf R}_{\bf y}^{\rm SDMA} &= \sigma_e^2 {\bf I} + \sum_{m'=1}^M (g_{q,m'}C_{{\rm sig},m'})^2\times \\
    &\left(\sum_{k'=1}^K {\bf H}_{m',k'}{\bf V}_{m',k'}{\bf V}_{m',k'}\H{\bf H}_{m',k'}\H + {\bf C}_{q,m'}\right).
    \end{aligned}
\end{equation}
Furthermore, the achievable SE of the $(m,k)$-th user is expressed as 
\begin{equation}
    {\rm SE}_{m,k}^{\rm SDMA} = \log_2\det\left({\bf I}_{N_r} + {\bf R}_{m,k}^{-1}{\bf H}_{m,k}{\bf V}_{m,k}{\bf V}_{m,k}\H {\bf H}_{m,k}\H\right). 
\end{equation}

We aim to maximize the weighted spectral efficiency\footnote{The spectral efficiency is defined as the total throughput [bps] divided by the total RF bandwidth [Hz]. } (WSE) ${\rm SE}_{\rm sum}$ of the multi-band uplink MU-MIMO quantum system by jointly optimizing the linear precoders ${\bf V}_{m,k}$, linear combiners ${\bf U}_{m,k}\H$, and the LO operating point $\{E_{{\rm LO},m}\}_{m=1}^M$. The optimization problem can be formulated as 
\begin{equation}
    \begin{aligned}
        \mathcal{P}_1:\quad \max_{{\bf V}_{m,k},{\bf a}_{\rm LO}}\quad &{\rm SE}_{\rm sum}^{\rm SDMA} = \frac{1}{M}\sum_{m=1}^M \sum_{k=1}^K \alpha_{m,k}{\rm SE}_{m,k}^{\rm SDMA}, \\ 
        \text{s.t.} ~~&\tr({\bf V}_{m,k}{\bf V}_{m,k}\H) \leq P_{m,k},~~\forall (m,k), \\ 
        & g_{q,m} = g_{q,m}({\bf a}_{\rm LO}),~~\forall m,
    \end{aligned} \label{eqn:problem_1}
\end{equation}
where $\alpha_{m,k}\geq 0$ is the weight of the $(m,k)$-th user, and ${\bf a}_{\rm LO}\in\mathbb{R}^{M\times 1}$ is the vector of LO configurations with the $m$-th entry being $E_{{\rm LO},m}$. The pre-factor $1/M$ is explained by the total RF bandwidth of $M\cdot{\rm BW}_{\rm IF}$ that appears on the denominator expression of SE. 
Unlike the traditional linear precoder and combiner design, this quantum variable ${\bf a}_{\rm LO}$ introduces additional degree of freedom of optimization to the uplink MU-MIMO Rydberg atomic receiver system.

\section{Algorithm Design} \label{sec3}
In this section, we solve the formulated quantum MU-MIMO uplink WSE maximization problem by iteratively optimizing the precoders $\{{\bf V}_{m,k}\}_{m,k=1}^{M,K}$, combiners $\{{\bf U}_{m,k}\H\}_{m,k=1}^{M,K}$, and the quantum-related variables ${\bf a}_{\rm LO}$.   

\subsection{Problem transformation}
We note that the objective function of the WSE maximization problem $\mathcal{P}_1$ is non-convex. To solve this problem, we first introduce two sets of auxiliary variables ${\bf U}_{m,k}, {\bf W}_{m,k}$ to transform the non-convex objective function ${\rm SE}_{\rm sum}^{\rm SDMA}({\bf V}_{m,k}, {\bf a}_{\rm LO})$ into a convex objective function $f_q({\bf U}_{m,k}, {\bf V}_{m,k}, {\bf W}_{m,k}, {\bf a}_{\rm LO})$. The new problem is 
\begin{equation}
\begin{aligned}
    \mathcal{P}_2:\quad \min_{{\bf U}_{m,k}, {\bf V}_{m,k}, {\bf W}_{m,k}, {\bf a}_{\rm LO}} f_q &= \frac{1}{\log 2}\sum_{m=1}^M\sum_{k=1}^K \alpha_{m,k}\times \\
    (\tr({\bf W}_{m,k}{\bf E}_{m,k})&-\log\det{\bf W}_{m,k}) \\ 
    \text{s.t.}~~ \tr({\bf V}_{m,k}{\bf V}_{m,k}\H)&\leq P_{m,k},~\forall (m,k),  \\
    g_{q,m} &= g_{q,m}({\bf a}_{\rm LO}),~\forall m,
\end{aligned}
\end{equation}
where ${\bf E}_{m,k}$ is a quadratic function of both ${\bf U}_{m,k}$ and $\{{\bf V}_{m,k}\}_{m=1,k=1}^{M,K}$, which is expressed as 
\begin{equation}
\begin{aligned}
    {\bf E}_{m,k} &= {\bf I}_S - g_{q,m}C_{{\rm sig},m} {\bf U}_{m,k}\H {\bf H}_{m,k}{\bf V}_{m,k} \\ 
    & -g_{q,m}C_{{\rm sig},m}{\bf V}_{m,k}\H {\bf H}_{m,k}\H{\bf U}_{m,k} \\ 
    & + {\bf U}_{m,k}\H {\bf R}_{\bf y}^{\rm SDMA} {\bf U}_{m,k}.
\end{aligned} \label{eqn:Emk}
\end{equation}
Note that the physical meaning of ${\bf E}_{m,k}$ is the MSE matrix of the linear detector $\hat{\bf s}_{m,k}^{\rm SDMA}$ in~\eqref{eqn:Linear-Detector}, which can be expressed as ${\bf E}_{m,k} = \mathbb{E}[(\hat{\bf s}_{m,k}^{\rm SDMA} - {\bf s}_{m,k})(\hat{\bf s}_{m,k}^{\rm SDMA} - {\bf s}_{m,k})\H]$. It has been proved that the optimal solution to $\mathcal{P}_2$ is equivalent to the optimal solution to $\mathcal{P}_1$, and vice versa~\cite[Theorem 1]{shi2011iteratively}. This equivalence also holds when the constraint of quantum transconductance $g_{q,m} = g_{q,m}({\bf a}_{\rm LO})$ is added to the optimization problem.

\subsection{Quantum WMMSE algorithm for Rydberg atomic receivers} 
In this subsection, we solve $\mathcal{P}_2$ by combining the renowned WMMSE algorithm~\cite{shi2011iteratively} with the quantum optimization techniques. The WMMSE algorithm solves the optimal ${\bf V}_{m,k}$ by applying the block coordinate descent method to alternately optimize ${\bf V}_{m,k}$ and the other two auxiliary variables ${\bf U}_{m,k}$ and ${\bf W}_{m,k}$. The update formula for ${\bf U}_{m,k}$ is given by the LMMSE detection formula as 
\begin{equation}
    {\bf U}_{m,k} = (g_{q,m}C_{{\rm sig},m})\left({\bf R}_{\bf y}^{\rm SDMA}\right)^{-1}{\bf H}_{m,k}{\bf V}_{m,k}, \label{eqn:update_U}
\end{equation}
and the update formula for ${\bf W}_{m,k}$ is given by the inverse of the MSE error matrix ${\bf E}_{m,k}$ as 
\begin{equation}
    {\bf W}_{m,k} = ({\bf I}_S-g_{q,m}C_{{\rm sig},m}{\bf U}_{m,k}\H {\bf H}_{m,k}{\bf V}_{m,k})^{-1}. \label{eqn:update_W}
\end{equation}
which can be obtained by plugging~\eqref{eqn:update_U} into~\eqref{eqn:Emk}. 
For the optimization of ${\bf V}_{m,k}$, we can fix ${\bf U}_{m,k}$ and ${\bf W}_{m,k}$, and treat the objective function $f_q$ as a quadratic function of ${\bf V}_{m,k}$. By dropping the  constant terms that are independent of ${\bf V}_{m,k}$, the objective function of $\mathcal{P}_2$ for each $(m,k)$ is simplified to be 
\begin{equation}
\begin{aligned}
    \mathcal{P}_3: \quad \min_{{\bf V}_{m,k}} f_{q,(m,k)}&=\frac{1}{\log 2}\sum_{m',k'} \alpha_{m',k'}\tr({\bf W}_{m',k'} {\bf E}_{m',k'}), \\ 
    \text{s.t.}\quad & \tr({\bf V}_{m,k}{\bf V}_{m,k}\H) \leq P_{m,k}. 
\end{aligned}
\end{equation}
where ${\bf E}_{m',k'}$ depends quadratically on $\{{\bf V}_{m,k}\}_{m=1,k=1}^{M,K}$. The problem $\mathcal{P}_3$ is a quadratically constrained quadratic programming (QCQP) problem, which can be solved using the Lagrange multiplier method. For each $(m,k)$, we introduce the Lagrange multiplier $\mu_{m,k}\geq 0$ and write the Lagrange function to be 
\begin{equation}
\begin{aligned}
    L({\bf V}_{m,k}, \mu_{m,k}) &= \sum_{m',k'}\alpha_{m',k'}\tr({\bf W}_{m',k'}{\bf E}_{m',k'}) \\
    &~+ \mu_{m,k}\tr({\bf V}_{m,k}{\bf V}_{m,k}\H). 
\end{aligned}
\end{equation}
By applying the KKT condition to the Lagrange function $L({\bf V}_{m,k}, \mu_{m,k})$, we obtain the first-order optimality condition of ${\bf V}_{m,k}$ to be 
\begin{equation}
\begin{aligned}
    {\bf V}_{m,k}^{\rm opt}(\mu_{m,k}) &= \left(\mu_{m,k}{\bf I}+(g_{q,m}C_{{\rm sig},m})^2{\bf H}_{m,k}\H{\bf F}{\bf H}_{m,k}\right)^{-1} \\ 
    &~\times \alpha_{m,k}g_{q,m}C_{{\rm sig},m}{\bf H}_{m,k}\H{\bf U}_{m,k}{\bf W}_{m,k},
\end{aligned} \label{eqn:update_V}
\end{equation}
where the matrix ${\bf F}\in\mathbb{C}^{N_r\times N_r}$ is defined as 
\begin{equation}
    {\bf F} = \sum_{m',k'}\alpha_{m',k'}{\bf U}_{m',k'}{\bf W}_{m',k'}{\bf U}_{m',k'}\H.
\end{equation}
The Lagrange multiplier $\mu_{m,k}^{\rm opt}$ is chosen to satisfy the complementary slackness condition $\mu_{m,k}\cdot(\tr({\bf V}_{m,k}{\bf V}_{m,k}\H)-P_{m,k}) = 0$, which can be obtained by a bisection search. 

The final step is to fix ${\bf U}_{m,k}, {\bf V}_{m,k}, {\bf W}_{m,k}$ and find the optimal ${\bf a}_{\rm LO}$ to minimize $f_q$. We notice that the objective function $f_q$ quadratically depends on $\{g_{q,m}\}_{m=1}^M$ via the MSE matrices ${\bf E}_{m,k}$. However, $\{g_{q,m}\}_{m=1}^M$ is generally not a convex function of ${\bf a}_{\rm LO}$, rendering the final step of quantum optimization non-convex. 

To solve this problem, we adopt a gradient-based optimizer for the variable ${\bf a}_{\rm LO}$, e.g., Armijo-Goldstein's backtrack search~\cite{dennis1996numerical}. The gradient $\partial f_q/\partial g_{q,\ell}$ is given by 
\begin{equation}
    \frac{\partial f_q}{\partial g_{q,\ell}} = \sum_{m,k}\alpha_{m,k}{\rm Tr}\left({\bf W}_{m,k}\frac{\partial {\bf E}_{m,k}}{\partial g_{q,\ell}}\right),~1\leq \ell\leq M, 
    \label{eqn:der_fq_gq}
\end{equation}
where 
\begin{equation}
\begin{aligned}
    \frac{\partial {\bf E}_{m,k}}{\partial g_{q,\ell}} &= -C_{{\rm sig},m}\delta_{m\ell}[{\bf U}_{m,k}\H{\bf H}_{m,k}{\bf V}_{m,k}+{\bf V}_{m,k}\H{\bf H}_{m,k}\H{\bf U}_{m,k}] \\ 
    &~+{\bf U}_{m,k}\H{\bf R}_\ell^\prime {\bf U}_{m,k}, 
\end{aligned}\label{eqn:der_Emk_gq}
\end{equation}
and
\begin{equation}
    {\bf R}_\ell^\prime = 2g_{q,\ell}C_{{\rm sig},\ell}^2\left(\sum_{k=1}^K{\bf H}_{\ell,k}{\bf V}_{\ell,k}{\bf V}_{\ell,k}\H{\bf H}_{\ell,k}\H + {\bf C}_{q,\ell}\right).\label{eqn:der_Rl_gq} 
\end{equation}
After solving $\partial f_q/\partial g_{q,\ell},\,\forall \ell$, the remaining step is to solve the quantum Jacobian ${\bf J}_q$ defined as $[{\bf J}_q]_{mn} = \partial g_{q,m}/\partial E_{{\rm LO},n}, \forall 1\leq m,n \leq M$. 

\subsection{Solving the quantum transconductance and quantum Jacobian}
The quantum transconductance $g_{q,m}$ is determined by $[\bar{\bm\rho}]_{21}$ via~\eqref{eqn:q_tcond}, which is further given by $\bar{\bf z}_0$ via~\eqref{eqn:rho_as_z}. Thus, we aim to evaluate $\partial \bar{\bf z}_0/\partial E_{{\rm LO},m},\,\forall m$, and $\partial^2\bar{\bf z}_0/\partial E_{{\rm LO},m}\partial E_{{\rm LO},n}, 1\leq m,n\leq M$. Since $\Omega_{{\rm LO},m} = \mu_{{\rm RF},m}E_{{\rm LO},m}/\hbar$, it is equivalent to evaluate $\partial \bar{\bf z}_0/\partial \Omega_{{\rm LO},m},\,\forall m$, and $\partial^2\bar{\bf z}_0/\partial \Omega_{{\rm LO},m}\partial \Omega_{{\rm LO},n}, 1\leq m,n\leq M$. For the sake of notational simplicity, we denote the derivative operator $\partial/\partial \Omega_{{\rm LO},m}$ as $\partial_m$. Apply $\partial_m$ to the equation ${\bf C}_0\bar{\bf z}_0 + {\bf w}_0/\sqrt{M+3}={\bf 0}$, we arrive at 
\begin{equation}
    \partial_m \bar{\bf z}_0 = -{\bf C}_0^{-1}\left((\partial_m {\bf C}_0)\bar{\bf z}_0 + \frac{1}{\sqrt{M+3}}\partial_m {\bf w}_0\right),\,\forall m.
\end{equation}
Furthermore, the second-order derivatives are computed by exploiting the linearity of ${\bf C}_0$ and ${\bf w}_0$ w.r.t. $\Omega_{{\rm RF},m}$, which are given by  
\begin{equation}
\begin{aligned}
    \partial_{m,n}^2 \bar{\bf z}_0 &= {\bf C}_0^{-1}\left[(\partial_n {\bf C}_0) {\bf C}_0^{-1}\left(\frac{\partial_m {\bf w}_0}{\sqrt{M+3}} + (\partial_m {\bf C}_0)\bar{\bf z}_0\right) \right.\\
    &\left. - (\partial_m {\bf C}_0)(\partial_n \bar{\bf z}_0) \right],\,\forall (m,n).
    \end{aligned}
\end{equation}
From the above computation, we can obtain $\partial_m [\bar{\bm \rho}]_{21}$ and $\partial_{mn}^2 [\bar{\bm \rho}]_{21}$ by applying~\eqref{eqn:rho_as_z}. Then, the quantum transconductance $g_{q,m}$ is computed as 
\begin{equation}
    g_{q,m} = \bar{I}_{\rm ph}\cdot \frac{2k_pN_0\mu_{12}^2}{\epsilon_0\hbar\Omega_p}(\partial_m [\bar{\bm \rho}]_{21})\cdot \frac{\mu_{{\rm RF},m}}{2\hbar}, 
    \label{eqn:update_gq}
\end{equation}
and the quantum Jacobian ${\bf J}_q$ is computed as 
\begin{equation}
\begin{aligned}
    [{\bf J}_q]_{mn} &= \bar{I}_{\rm ph}\frac{2k_pN_0\mu_{12}^2}{\epsilon_0\hbar\Omega_p}(\partial_{mn}^2[\bar{\bm \rho}]_{21}))\frac{\mu_{{\rm RF},m}\mu_{{\rm RF},n}}{(2\hbar)^2} \\ 
    +&\bar{I}_{\rm ph}L\left(\frac{2k_pN_0\mu_{12}^2}{\epsilon_0\hbar\Omega_p}\right)^2(\partial_m [\bar{\bm \rho}]_{21})(\partial_n [\bar{\bm \rho}]_{21})\frac{\mu_{{\rm RF},m}\mu_{{\rm RF},n}}{(2\hbar)^2}. 
\end{aligned}\label{eqn:update_Jq}
\end{equation}

Combining the precoder/combiner updating rules and the evaluation of quantum transconductance and quantum Jacobian, the proposed qWMMSE algorithm is summarized in~{\bf Algorithm~\ref{alg:qWMMSE}}. 

\begin{algorithm}[!t] 
    \caption{Proposed qWMMSE algorithm} \label{alg:qWMMSE}
    \setstretch{1.2}
    \begin{algorithmic}[1]
	\REQUIRE
		Uplink channels $\{{\bf H}_{m,k}\}_{m,k=1}^{M,K}$, power constraints $\{P_{m,k}\}_{m,k=1}^{M,K}$, user weights $\{\alpha_{m,k}\}_{m,k=1}^K$, BBR covariance matrices $\{{\bf C}_{q,m}\}_{m=1}^M$, electronic noise variance $\sigma_e^2$, precision threshold $\epsilon$. 
	\ENSURE 
		Optimal precoders $\{{\bf V}_{m,k}\}_{m,k=1}^{M,K}$, combiners $\{{\bf U}_{m,k}\H\}_{m,k=1}^{M,K}$, LO field strengths ${\bf a}_{\rm LO}$. 
        \vspace{3pt}
        \STATE Randomly initialize ${\bf V}_{m,k}$ to satisfy $\tr({\bf V}_{m,k}{\bf V}_{m,k}\H) = P_{m,k},\,\forall(m,k)$. 
        \REPEAT
            \STATE ${\bf W}_{m,k}' \leftarrow {\bf W}_{m,k}$; 
            \STATE Update $\{g_{q,m}\}_{m=1}^M$ and ${\bf J}_q$ with \eqref{eqn:update_gq} and \eqref{eqn:update_Jq}; 
            \STATE Update ${\bf U}_{m,k}\H$ with~\eqref{eqn:update_U}, $\forall (m,k)$;  
            \STATE Update ${\bf W}_{m,k}$ with~\eqref{eqn:update_W}, $\forall (m,k)$; 
            \STATE Update ${\bf V}_{m,k}$ with~\eqref{eqn:update_V}, $\forall (m,k)$; 
            \STATE Update the derivatives $\partial f_q/\partial {g}_{q,m},\forall m$ with \eqref{eqn:der_fq_gq},\eqref{eqn:der_Emk_gq},\eqref{eqn:der_Rl_gq}; 
            
            \STATE Update ${\bf a}_{\rm LO}$ with the gradient $\partial f_q/\partial {\bf a}_{\rm LO}$ and Armijo-Goldstein backtrack search to minimize $f_q$; 
        \UNTIL{$\sum_{m,k}\left|\log\det {\bf W}_{m,k} - \log\det{\bf W}_{m,k}'\right| \leq \epsilon$}. 
        \RETURN $\{{\bf V}_{m,k}\}_{m,k=1}^{M,K}$, $\{{\bf U}_{m,k}\H\}_{m,k=1}^{M,K}$, ${\bf a}_{\rm LO}$. 
    \end{algorithmic}
\end{algorithm}

\subsection{Algorithm complexity}
Let $N_{\rm iter}$ be the number of qWWMSE iterations. 
Since the computation of quantum transconductances $\{g_{q,m}\}_{m=1}^M$ requires $M$ matrix inversion operations of size $\mathcal{O}(M^2)$, this step consumes $\mathcal{O}(M^7)$ floating point operations (FLOPs). Furthermore, the computation of quantum Jacobian ${\bf J}_q$ consumes $\mathcal{O}(M^8)$ FLOPs. 

The computation of ${\bf U}_{m,k}\H$ consumes $\mathcal{O}(N_r^3+MK(N_tN_rS+N_r^2S))$ FLOPs. 
The computation of ${\bf W}_{m,k}$ consumes $\mathcal{O}(MK(S^3+N_tN_rS))$ FLOPs. 
The computation of ${\bf V}_{m,k}$ consumes $\mathcal{O}(MK(N_tS^2+ (N_r^2N_t+N_t^3+N_tN_rS+N_rS^2)\log(1/\epsilon))$ FLOPs to achieve an error below $\epsilon$. 
The computation of $\partial f_q/\partial g_{q,m}, \forall m$ consumes $\mathcal{O}(M^2K(N_rN_tS+N_r^2S)+MKN_tS^2)$ FLOPs. 

Finally, the computation of $\partial f_q/\partial {\bf a}_{\rm LO}$ consumes $\mathcal{O}(M^2)$ FLOPs. Generally speaking, if we assume that $N_r$ is of the same order as $N_t$, then the computational complexity scales as $\mathcal{O}(N_{\rm iter}M^8+N_{\rm iter}MKN_r^2S\log(1/\epsilon))$, which is a cubic growth with the number of BS Rx vapor cells. Although the $\mathcal{O}(M^8)$ growth scales very fast with the number of RF bands $M$, we point out that $M$ will not exceed 5 for practical simultaneous atomic RF reception~\cite{allinson2024simultaneous}.

\section{Numerical Results} \label{sec4}
In this section, we present the numerical results of the proposed qWMMSE algorithm. 

\subsection{Experimental setup}
In the following numerical computation, we assume that the BS is equipped with $N_r=5$ atomic vapor cells filled with Cesium-133 vapor with atomic density $N_0=4.89\times 10^{10}\,{\rm cm}^{-3}$. We consider a dual-band ($M=2$) quantum receiver with the Rydberg atoms described by a five-level quantum system. The energy levels $\{\ket{n}\}_{n=1}^5$ are specified as $\ket{1}=\ket{6S_{1/2}}$, $\ket{2}=\ket{6P_{3/2}}$, $\ket{3}=\ket{47D_{5/2}}$, $\ket{4}=\ket{48P_{3/2}}$, and $\ket{5}=\ket{45F_{7/2}}$. The center transition frequencies are $f_{c,1}=f_{\ket{3}\leftrightarrow\ket{4}}=6.938\,{\rm GHz}$  $f_{c,2}=f_{\ket{3}\leftrightarrow\ket{5}} = 31.793\,{\rm GHz}$, which is also the frequencies to which the atomic vapor is sensitive. The Rabi frequencies and detunings\footnote{These detuning values are set to small non-zero values to reflect the practical frequency drifts of laser sources and RF LO sources. } of the probe, control, LO1, and LO2 fields are set to be $\Omega_p=2\pi\times 8.08\,{\rm MHz}$, $\Omega_c=2\pi\times 2.05\,{\rm MHz}$, $\Delta_p=2\pi\times 20\,{\rm Hz}$, $\Delta_c=2\pi\times (-30)\,{\rm Hz}$, $\Delta_{\ell,1}=2\pi\times 10\,{\rm Hz}$, and $\Delta_{\ell,2}=2\pi\times 20\,{\rm Hz}$. The length of the vapor cell is set to be $L=2\,{\rm cm}$ with an ambient temperature of $T=300\,{\rm K}$. The decay rates of the quantum states $\{\ket{n}\}_{n=2}^5$ are set to be $\gamma_2=2\pi\times 5.2\,{\rm MHz}$, $\gamma_3=2\pi\times 3.9\,{\rm kHz}$, $\gamma_4=2\pi\times 1.7\,{\rm kHz}$, and $\gamma_5=2\pi\times 1.6\,{\rm kHz}$. For the electronic transimpedance (TIA) amplifier that amplifies the photocurrent, the transimpedance is $R_{\rm T}=10\,{\rm k\Omega}$, the input impedance is $Z_{\rm in}=60\,\Omega$, the output impedance is $Z_{\rm out} = 50\,{\Omega}$, the PD bias resistor is $R_s=1\,{\rm k\Omega}$, the input-referred voltage noise is $V_n=2.8\,{\rm nV}/\sqrt{\rm Hz}$, and the input-referred current noise is $I_{n}=1.8\,{\rm pA/\sqrt{Hz}}$~\cite{DHPCA100}. The reference voltage is $V_{\rm ref}=1\,{\rm mV}$~\cite{zhu2025general}. The IF bandwidth is set to be ${\rm BW}_{\rm IF}=100\,{\rm kHz}$~\cite{du2022realization,zhu2025general}. 

For MU-MIMO uplink communications, the classical users are equipped with $N_t=4$ Tx antennas. For each of the $M=2$ bands, the number of users is set to be $K=3$. The large-scale path loss is determined by the user's distance, which is sampled from $\mathcal{U}(500\,{\rm m}, 1500\,{\rm m})$. The small-scale fading is modeled by i.i.d. Rayleigh fading. The minimum allowed LO E-field intensity is $E_{{\rm LO}}\geq 3\,{\rm mV/m}$.

\subsection{Baseline schemes}
To demonstrate the benefits of the quantum optimization algorithm for RAQRs, we choose the baseline schemes to be the standard WMMSE precoding without quantum optimization (No-Opt) for quantum receivers and the WMMSE algorithm for classical receivers. The baselines are detailed as follows. 
\begin{enumerate}
    \item {\bf cSDMA with MC}. A classical receiver array with $N_R$ antennas is installed at the BS to simultaneously serve $MK$ uplink users, where the antenna mutual coupling (MC) is considered. 
    \item {\bf cSDMA without MC}. This scheme differs from the cSDMA with MC scheme by not considering the MC effects. 
    \item {\bf qSDMA-Opt}. A Rydberg atomic receiver array with $N_R$ vapor cells is installed at the BS to simultaneously serve $K$ uplink users in each of the $M$ RF bands, where the quantum optimization of ${\bf a}_{\rm LO}$ is enabled. All of the $MK$ users share the entire IF bandwidth, and they are distinguished by the Rx spatial filter ${\bf U}_{m,k}\H$. 
    \item {\bf qFDMA-Opt}. A Rydberg atomic receiver array with $N_R$ vapor cells is installed at the BS to simultaneously serve $K$ uplink users in each of the $M$ bands, where the quantum optimization of ${\bf a}_{\rm LO}$ is enabled. The entire IF band is evenly divided into $M$ sub-bands, where user signals from different RF bands are down-converted to occupy different parts of the IF band. Users in the same RF band are distinguished by the Rx spatial filter ${\bf U}_{m,k}\H$. 
    \item {\bf qSDMA-NoOpt}. This scheme differs from the qSDMA-Opt scheme by disabling the quantum optimization of ${\bf a}_{\rm LO}$. 
    \item {\bf qFDMA-NoOpt}. This scheme differs from the qFDMA-Opt scheme by disabling the quantum optimization of ${\bf a}_{\rm LO}$. 
\end{enumerate}

For the qFDMA schemes, the signal model slightly differs from~\eqref{eqn:MU-MIMO_signal_model} by processing each of the $M$ bands separately. The FDMA signal model is given by 
\begin{equation}
    {\bf y}_m^{\rm FDMA} = g_{q,m} C_{{\rm sig},m} \left(\sum_{k=1}^K {\bf H}_{m,k}{\bf V}_{m,k}{\bf s}_{m,k}\right) + {\bf w}_{{\rm tot},m}',\,\forall m, \label{eqn:FDMA_signal_model}
\end{equation}
where the covariance matrix of ${\bf w}_{{\rm tot},m}'$ is ${\bf C}_{\rm tot}/M$ due to the divided IF band. The WSE expression of the qFDMA scheme is the same as in~\eqref{eqn:problem_1}. 
Note that the signal model of the qFDMA scheme is generally aligned with~\cite{cui2025rydberg}, while the noise model ${\bf C}_{\rm tot}/M$ is different from~\cite{cui2025rydberg} by carefully considering the in-band/out-of-band BBR noise, the image frequency BBR noise, and the noise superposition effect in the IF band. 

The qWMMSE algorithm in the FDMA scheme is similar to that of the SDMA scheme, but the updating formulas for ${\bf U}_{m,k}, {\bf V}_{m,k}, {\bf W}_{m,k}$, and ${\bf a}_{\rm LO}$ are modified according to the FDMA signal model~\eqref{eqn:FDMA_signal_model}. See {\bf Appendix~\ref{app:FDMA_updating_formulas}} for details.

\subsection{Performance of qWMMSE}
\begin{figure}
    \centering
    \includegraphics[width=\figsize\linewidth]{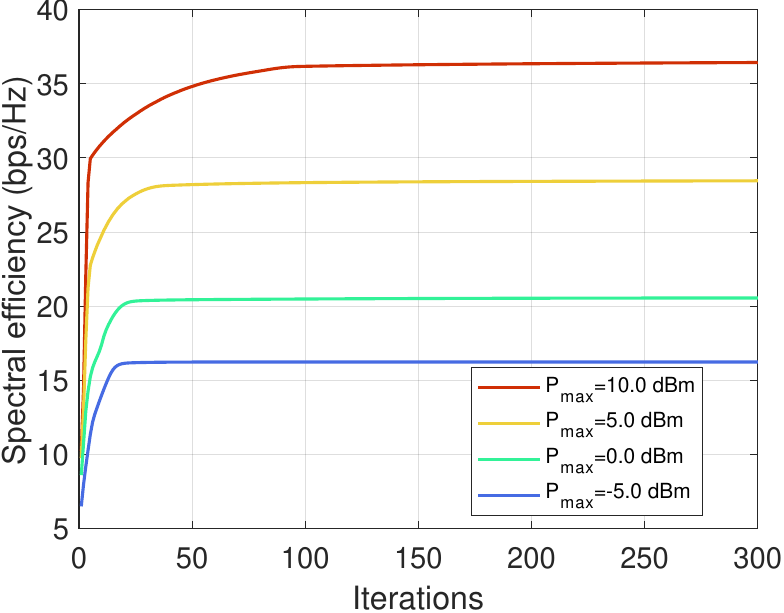}
    \caption{Weighted SE v.s. number of iterations, evaluated with different uplink transmission power. }
    \label{fig:Technical_wRate_Iters}
\end{figure}

To show the convergence of the proposed qWMMSE {\bf Algorithm~\ref{alg:qWMMSE}}, we present the optimization trajectory as a function of qWMMSE iterations in Fig.~\ref{fig:Technical_wRate_Iters}. 

\begin{figure}
    \centering
    \includegraphics[width=\figsize\linewidth]{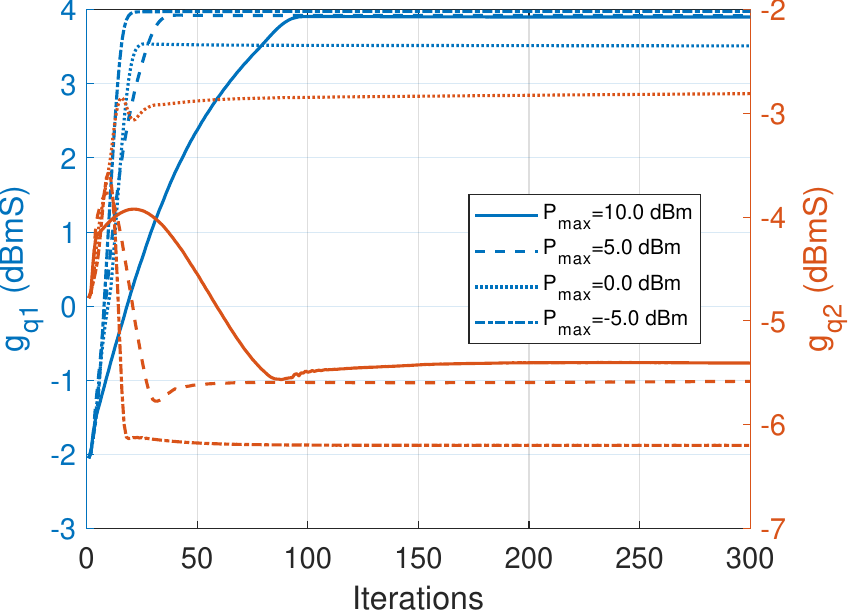}
    \caption{Quantum transconductance $g_{q,m}$ v.s. number of iterations, evaluated with different uplink transmission power. }
    \label{fig:Technical_gqs_Iters}
\end{figure}

Fig.~\ref{fig:Technical_wRate_Iters} shows the improvement in ${\rm SE}_{\rm sum}^{\rm SDMA}$ w.r.t. number of iterations. The SE improves monotonically as the iteration proceeds, demonstrating that each alternating optimization step can reduce the objective function $f_q$. For a low maximum transmit power budget $P_{\rm max}$, the algorithm convergence is faster, which is explained by the relatively small number of active users in this power-restricted regime. Fig.~\ref{fig:Technical_gqs_Iters} presents the convergence of quantum transconductance values $g_{q,1}$ and $g_{q,2}$ w.r.t. number of iterations, which exhibits similar convergence behavior to that of SE. 

\begin{figure}
    \centering
    \includegraphics[width=\figsize\linewidth]{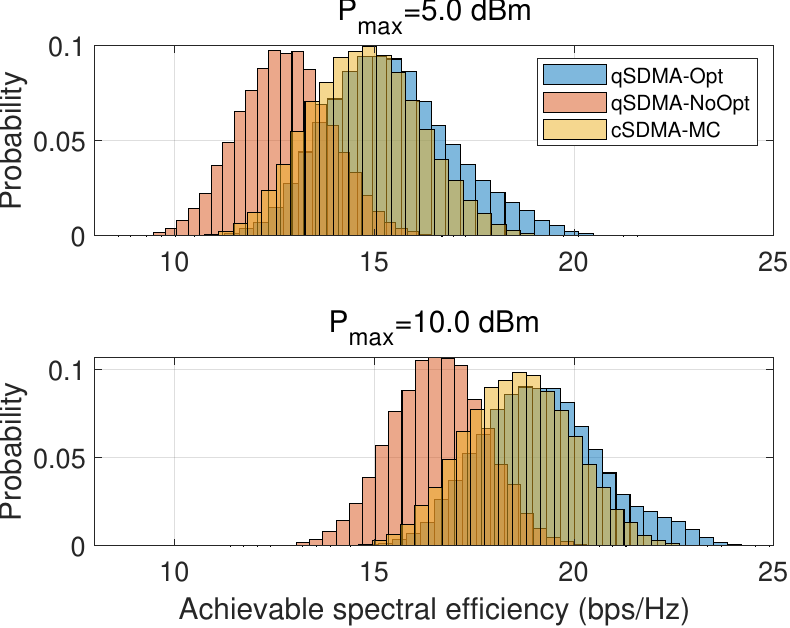}
    \caption{Distribution of the weighted SE achieved by different MU-MIMO schemes, evaluated from 30,000 Monte Carlo trials.  }
    \label{fig:Technical_SE_Hist}
\end{figure}

Fig.~\ref{fig:Technical_SE_Hist} shows the distribution of the achievable SE of different MU-MIMO SDMA schemes. It can be observed that, the average rate performance of the proposed qSDMA-Opt scheme is higher than that of the classical SDMA schemes and that of the qSDMA-NoOpt scheme. However, the rate variance of the qSDMA-Opt scheme is slightly larger. 

\begin{figure}
    \centering
    \includegraphics[width=\figsize\linewidth]{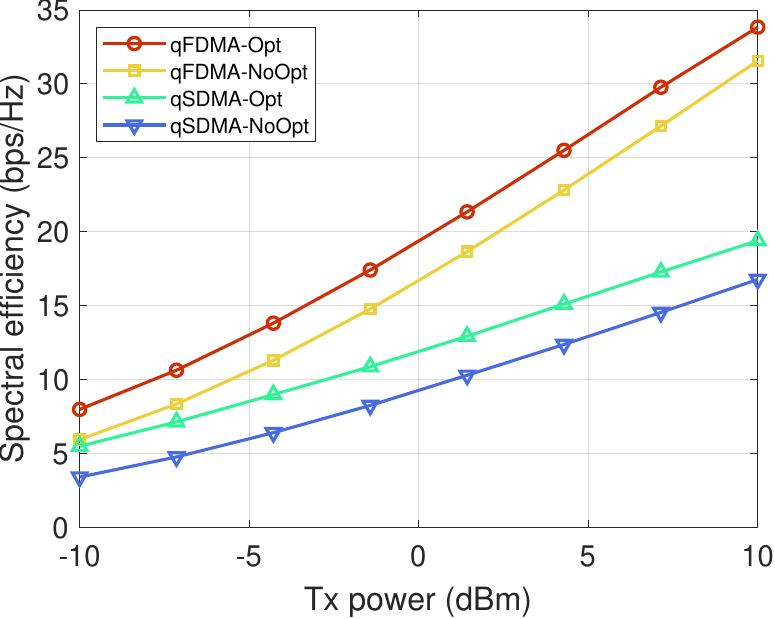}
    \caption{Weighted spectral efficiency v.s. transmit power constraint. }
    \label{fig:Technical_qOpt}
\end{figure}

To analyze the performance of the proposed qWMMSE algorithm against all the baselines, we tested the MU-MIMO quantum reception schemes with different Tx transmit power. Fig.~\ref{fig:Technical_qOpt} shows the SE performance of different multi-band quantum reception schemes as a function of the Tx power budget. Compared with the NoOpt schemes, the proposed quantum optimization procedure of ${\bf a}_{\rm LO}$ uniformly improves the overall spectral efficiency by approximately $3\,{\rm bps/Hz}$. The qFDMA schemes generally outperform the qSDMA schemes in terms of SE, which is mainly due to the $M$-fold RF bandwidth reduction of the qFDMA users compared to that of the qSDMA users. 

\begin{figure}
    \centering
    \includegraphics[width=\figsize\linewidth]{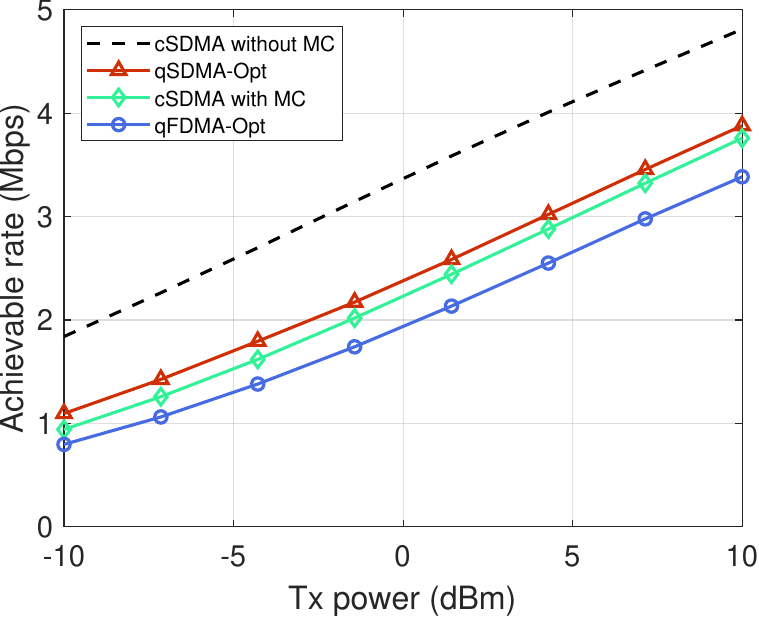}
    \caption{Achievable sum rate v.s. transmit power constraint. }
    \label{fig:Technical_CvsQ}
\end{figure}

Fig.~\ref{fig:Technical_CvsQ} compares the achievable sum rate of quantum receivers with classical electronic receivers of the same IF bandwidth. Despite the additional noises of quantum receivers (see {\bf Table~\ref{tab:Noise_Sources}}), the multi-band qSDMA receiver with quantum optimization procedure could possibly outperform the classical receivers with antenna mutual coupling~\cite{yuan2023rydberg}. This benefit of quantum receivers is due to the weakened electromagnetic mutual coupling between the optically accessed atomic vapor cells~\cite{gong2025rydberg}. Moreover, the qSDMA schemes outperform qFDMA in terms of achievable sum rate, which is mainly due to the $M$-fold enlarged RF bandwidth. 

\section{Conclusions} \label{sec5}
In this paper, we proposed the multi-band Rydberg atomic quantum MIMO architecture (RAQ-MIMO) to overcome the intermediate frequency interference (IFI) problem of multi-band Rydberg atomic receivers. 
The signal model was characterized by a quantum transconductance-based physically compliant model, together with a noise model considering various noise sources. 
The SE optimization problem of RAQ-MIMO systems was then solved by the qWMMSE algorithm, where joint optimization of the classical linear precoders/combiners and the quantum LO configurations was performed. 
Simulation results have demonstrated the convergence and effectiveness of the proposed qWMMSE algorithm. 

Future works will be focused on the extremely wideband RF reception~\cite{djoumessi2009frequency} enabled by the multi-band property of Rydberg atoms, and the experimental validation of the RAQ-MIMO signal model.

\section{Acknowledgment}
The authors would like to thank Prof. Wei E. I. Sha from Zhejiang University and Dr. Hanfeng Wang from Massachusetts Institute of Technology for their helpful discussions and constructive suggestions on Rydberg atomic receivers.  

\appendices 

\section{Computation of the BBR noise covariance} \label{app:BBR_cov_formula}
Let us consider an ambient blackbody radiation field of temperature $T\,{\rm [K]}$. By Planck's formula, the BBR field energy $u_\nu\,{\rm [J/m^{3}/Hz]}$ per unit volume per unit frequency is given by 
\begin{equation}
    u_\nu(T) = \frac{4\pi}{c_0}B_\nu(T), \label{eqn:BBR_energy_density}
\end{equation}
where $B_\nu(T)$ is the BBR spectral radiance expressed as 
\begin{equation}
    B_\nu(T) = \frac{2\nu^2}{c_0^2}\frac{h\nu}{e^{h\nu/k_{\rm B}T}-1}\approx \frac{2\nu^2}{c_0^2}k_{\rm B}T,
\end{equation}
and $\nu\,{\rm [Hz]}$ is the center frequency at which the BBR is measured. Let $E_{{\rm bbr},z}\,{\rm [V/m/\sqrt{Hz}]}$ be the BBR noise phasor per unit rooted Hertz along the $z$-axis, which is a circularly Gaussian distributed random variable. According to the energy equipartition theorem at thermal equilibrium, the unit-volume BBR electric field energy $\epsilon_0|E_{\rm bbr}|^2/4$ and the unit-volume BBR magnetic field energy are equal. Thus, we have 
\begin{equation}
    2\times \frac{3}{4}\epsilon_0\mathbb{E}\left[|E_{{\rm bbr},z}|^2\right] = u_\nu(T). 
\end{equation}
Combining with the BBR energy density formula~\eqref{eqn:BBR_energy_density}, we have 
\begin{equation}
    \mathbb{E}\left[|E_{{\rm bbr},z}|^2\right] = \frac{16\pi\eta_0}{3\lambda_0^2}k_{\rm B}T. 
    \label{eqn:variance_bbr_efield}
\end{equation}
Note that the coefficient $\sqrt{2\eta_0/A_e}$ ($A_e=\lambda_{c,m}^2/(4\pi)$) that appears in the definition of the quantum reference power~\eqref{eqn:quantum_reference_power} is for the conversion of the incident rooted signal power $\rm [\sqrt{W}]$ to the signal E-field $E_{\rm sig}\,{\rm [V/m]}$. Divide~\eqref{eqn:variance_bbr_efield} by the square of this coefficient, we get $2k_{\rm B}T/3$, which is the variance of the equivalent incidence BBR noise measured in $\rm [W/Hz]$. Finally, the formula of~\eqref{eqn:quantum_noise_components} is proved by considering the factor of 2 brought by the image frequency BBR noise, the BBR coherence factor $\zeta(\ell)$~\cite{zhu2025general}, and the normalized BBR noise correlation described by $\hat{\bf C}_{q,m}$ in each band $m$.

\section{qWMMSE updating formulas for FDMA} \label{app:FDMA_updating_formulas}
In the FDMA signal model~\eqref{eqn:FDMA_signal_model}, signals from each of the $M$ bands are received separately. The updating formula for ${\bf U}_{m,k}$ is given by 
\begin{equation}
    {\bf U}_{m,k} = (g_{q,m}C_{{\rm sig},m}){\bf R}_m^{-1} {\bf H}_{m,k}{\bf V}_{m,k}, 
\end{equation}
where the signal covariance matrix of the $m$-th band ${\bf R}_{m}$ is defined as 
\begin{equation}
\begin{aligned}
    {\bf R}_m &:= \frac{1}{M}{\bf C}_{\rm tot}  \\ 
    &~+ (g_{q,m}C_{{\rm sig},m})^2\sum_{k'=1}^K {\bf H}_{m,k'}{\bf V}_{m,k'}{\bf V}_{m,k'}\H{\bf H}_{m,k'}\H. 
\end{aligned}
\end{equation}
The updating formula for ${\bf W}_{m,k}$ is given by 
\begin{equation}
    {\bf W}_{m,k} = \left({\bf I}_S - g_{q,m}C_{{\rm sig},m}{\bf U}_{m,k}\H{\bf H}_{m,k}{\bf V}_{m,k} \right)^{-1}, 
\end{equation}
and the MMSE error matrix ${\bf E}_{m,k}$ is expressed as 
\begin{equation}
\begin{aligned}
    {\bf E}_{m,k} &= {\bf I} - g_{q,m}C_{{\rm sig},m} {\bf U}_{m,k}\H {\bf H}_{m,k}{\bf V}_{m,k}  \\
    &~- g_{q,m}C_{{\rm sig},m}{\bf V}_{m,k}\H {\bf H}_{m,k}\H {\bf U}_{m,k} + {\bf U}_{m,k}\H{\bf R}_m{\bf U}_{m,k}. 
\end{aligned}
\end{equation}
The updating formula for ${\bf V}_{m,k}$ is given by 
\begin{equation}
\begin{aligned}
    {\bf V}_{m,k}(\mu_{m,k}) &= \alpha_{m,k}g_{q,m}C_{{\rm sig},m}\left(\mu_{m,k}{\bf I} + {\bf H}_{m,k}\H {\bf A}_m {\bf H}_{m,k}\right)^{-1} \\ 
    & \times {\bf H}_{m,k}\H {\bf U}_{m,k}{\bf W}_{m,k}, 
\end{aligned}
\end{equation}
where the Lagrange multiplier $\mu_{m,k}\geq 0$ can be obtained by bisection search, and the matrix ${\bf F}_m\in\mathbb{C}^{N_r\times N_r}$ is defined as 
\begin{equation}
    {\bf F}_m = \sum_{k'=1}^K \alpha_{m,k'}{\bf U}_{m,k'}{\bf W}_{m,k'}{\bf U}_{m,k'}\H,\,\forall m. 
\end{equation}
For the Armijo-Goldstein update formula of the quantum parameters, the derivative ${\bf R}_m'$ in~\eqref{eqn:der_Rl_gq} is replaced by 
\begin{equation}
    \frac{\partial {\bf R}_m}{\partial g_{q,\ell}} = 2g_{q,\ell}C_{{\rm sig},\ell}^2 \left(\delta_{m\ell}\sum_{k=1}^K {\bf H}_{\ell,k}{\bf V}_{\ell,k}{\bf V}_{\ell,k}\H{\bf H}_{\ell,k}\H + \frac{1}{M}{\bf C}_{q,\ell}\right). 
\end{equation}

\footnotesize

\bibliographystyle{IEEEtran}
\bibliography{IEEEabrv, bibfile}

\end{document}